\documentclass[preprint,preprintnumbers,amsmath,amssymb]{revtex4-1}
\usepackage{amsfonts}    
\usepackage{amssymb}
\usepackage{amsmath}
\usepackage{latexsym}
\usepackage{eepic}
\usepackage{graphicx}
\usepackage[usenames,dvipsnames]{color}
\usepackage{color}

\usepackage[active]{srcltx}
\usepackage{epstopdf}

\newcommand{\pa}{\partial}

\newcommand{\om}{\omega}

\newcommand{\De}{\Delta}

\begin{document}

\title{On the four-body problem in the Born-Oppenheimer approximation}

\author{C. A. Escobar}
\email{carlos\_escobar@fisica.unam.mx}
\affiliation{Instituto de F\'{i}sica, Universidad Nacional Aut\'{o}noma de M\'{e}xico, Apartado Postal 20-364, Ciudad de M\'{e}xico 01000, M\'{e}xico}

\author{A. Mart\'{i}n-Ruiz}
\email{alberto.martin@nucleares.unam.mx}
\affiliation{Instituto de Ciencias Nucleares, Universidad Nacional Aut\'{o}noma de M\'{e}xico, 04510 Ciudad de M\'{e}xico, M\'{e}xico}

\begin{abstract}

The quantum problem of four particles in $\mathbb{R}^d$ ($d\geq 3$), with arbitrary masses $m_1,m_2,m_3$ and $m_4$, interacting through an harmonic oscillator potential is considered. This model allows exact solvability and a critical analysis of the Born-Oppenheimer approximation. The study is restricted to the ground state level. We pay special attention to the case of two equally heavy masses $m_1=m_2=M$ and two light particles $m_3=m_4=m$. It is shown that the sum of the first two terms of the Puiseux series, in powers of the dimensionless parameter $\sigma=\frac{m}{M}$, of the exact phase $\Phi$ of the wave function $\psi_0=e^{-\Phi}$ and the corresponding ground state energy $E_0$, coincide exactly with the values obtained in the Born-Oppenheimer approximation. A physically relevant rough model of the $H_2$ molecule and of the chemical compound $H_2O_2$ (Hydrogen peroxide) is described in detail. The generalization to an arbitrary number of particles $n$, with $d$ degrees of freedom ($d\geq n-1$), interacting through an harmonic oscillator potential is briefly discussed as well.

\end{abstract}

\maketitle

\newpage

\section{Introduction }

Recently, a many-body quantum reduced Hamiltonian was presented in Ref. \cite{TME-N}. It describes the ground state level of a $d$-dimensional $n$-body quantum system, $d\geq n-1$ , with an arbitrary potential that solely depends on relative distances between the particles. The dynamical variables of the reduced Hamiltonian are the $\frac{n(n-1)}{2}$ relatives distances between the bodies. In particular, in Refs. \cite{TME3-d}-\cite{TME3-d2} the three-body system was considered in detail while in the work \cite{TME4-d} the four-body case was analyzed. Also, for the four-body quantum system Jacobi-like-variables were introduced in Ref. \cite{Gu,extra1} to reduce the Schr\"{o}dinger equation to generalized radial equations where only six internal variables are involved. It is worth mentioning that even in the planar case $d=2$, the dynamics of the classical 4-body problem is very rich \cite{Erdi}-\cite{Albouy}. Unfortunately, in all these important conceptual works few applications are mentioned and not explicit examples were worked out in detail.

The Born-Oppenheimer approximation (BOA) \cite{Born1,Born2} is a landmark in atomic and molecular physics. Even the well-established notion of molecular electronic states is deeply connected to this approximation.

The general assumption of this approximation is physically transparent: when the masses of the nuclei are much heavier than the electronic ones then an approximate separation of the electronic and nuclear motions (pseudo-separation of variables) is possible and, consequently, the task of solving the Schr\"{o}dinger equation or the classical Hamilton's equations become considerably simpler.

The nature of the BOA is essentially perturbative. In the case of a quantum Coulomb system of electrons and nuclei, the small expansion parameter $\lambda=\sigma^{\frac{1}{4}}$ involves the ratio of the electron mass $m$ to the nuclear mass $M$, namely $\sigma = m/M $ \cite{Born1,Born2}. Accordingly, the Hamiltonian is expressed as the sum of two terms, the so-called clamped-nuclei electronic Hamiltonian (which is $\sigma$-independent) and the nuclear kinetic energy operator ($\propto \,\sigma$). As a result, in the BOA the total approximate molecular wavefunction is factorized as a product of an electronic and a nuclear wavefunction. In fact, Hunter \cite{Hunter} suggested that the exact wavefunction also admits such a factorization. Remarkably, using a variational principle Cederbaum showed that this factorization occurs indeed \cite{Cederbaum}.

In order to examine the accuracy of the Born-Oppenheimer approximation, Moshinsky and Kittel \cite{Moshinski} discussed the 3-dimensional ($d=3$) elementary problem of a light particle (the electron) and two heavy particles (the nuclei) which are coupled to each other by harmonic
forces. This model, originally used in nuclear physics \cite{Green}, can be solved both exactly and within BOA, thus allowing a critical discussion of such an approximation. The analysis in Ref. \cite{Moshinski} concluded that the Born-Oppenheimer approximation provides accurate results for both the energy and the ground-state wave function, even for the extreme case in which the \emph{light} particle is a proton in a hydrogen bond.

In a recent paper, Sutcliffe and Woolley made a careful reformulation of the conventional Born-Oppenheimer argument drawing on results from the modern mathematical literature and proved that the correct $\sigma$-independent part of Hamiltonian is not the clamped-nuclei electronic Hamiltonian but a different operator which was given there explicitly. Until now, due to its enormous importance in contemporary experimental atomic and molecular physics, the BOA continues to be a fruitful object of study and it is used in a plethora of applications.

The purpose of the present work is two fold. On the one hand, for a 3-body ($n=3$) exactly solvable model in $\mathbb{R}^3$ ($d=3$), based on the formalism described in Ref. \cite{TME3-d}, we will re-derive the known results \cite{Moshinski} in a cleaner and more elegant manner. Afterwards, we extend these 3D results to the case of an arbitrary dimension $d>3$. Secondly, along the same lines we aim to determine quantitatively the accuracy of the Born-Oppenheimer approximation for a 4-body closed chain of interacting harmonic oscillators with $d$-degrees of freedom ($d>2$). At $d=3$, such a system is a rough model of the $H_2$ molecule and of the chemical compound $H_2O_2$ (Hydrogen peroxide). Valuable results for the $n$-body case in $d$-dimensions are derived as well.

\section{Three body problem}

In this Section we first consider a three-body quantum system in $d$-dimensions ($d>1$). The Hamiltonian is of the form
\begin{equation}
\label{Hgen}
   {\cal H}\ =\ -\sum_{i=1}^3 \frac{1}{2\, m_i} \De_i^{(d)}\ +\  V(r_{12},\,r_{13},\,r_{23})\ ,\
\end{equation}
where $V$ is a scalar potential that solely depends on the relative distances
\[
r_{12}\ = \ \mid {\bf r}_{1}-{\bf r}_{2}\mid \qquad , \quad r_{13}\ = \ \mid {\bf r}_{1}-{\bf r}_{3}\mid \qquad , \quad r_{23} \ = \ \mid {\bf r}_{2}-{\bf r}_{3}\mid \ ,
\]
and $\De_i$ is the individual $d$-dimensional Laplacian
\[
     \De_i \ =\ \frac{\pa^2}{\pa{{\bf r}_i} \pa{{\bf r}_i}}\ ,
\]
associated to the $i$th body with mass $m_i$ and coordinate vector ${\bf r}_i =(x_{i,1}\,, x_{i,3}\ldots \,,x_{i,d})\in \mathbb{R}^d$\,. Thus, in total the Hamiltonian (\ref{Hgen}) possesses $3d$ degrees of freedom. After separating the center of mass motion, the number of degrees of freedom of the system reduces to $2d$ in the space of relative motion.

For this reduced system there exists a quadratic potential $V$ in terms of relative distances $r_{12},\,r_{23},\,r_{13}$, for which uncountable number of quantum $S$-states (zero total angular momentum) of the eigenvalue problem ${\cal H}\,\Psi=E\,\Psi$ can be found by algebraic means.
Their eigenfunctions are the elements of the finite-dimensional representation space(s) of $\mathfrak{sl}(4,{\bf R})$ algebra of differential operators \cite{TME3-d}. In general, assuming that the potential
\begin{equation}\label{Vpo}
V\ = \ V(\,\rho_{12},\,\rho_{13},\,\rho_{23}\,) \ ,
\end{equation}
solely depends on the $\rho$-variables

\begin{equation}\label{redi}
\rho_{12}\ = \ r_{12}^2\ ,\qquad  \rho_{13}\ = \ r_{13}^2\ ,\qquad \rho_{23} \ = \ r_{23}^2\ ,
\end{equation}

(relative distances squared) we arrive at the reduced three-dimensional radial equation in the $\rho$-space \cite{TME3-d}
\begin{equation}
\label{Hrad3}
       {\cal H}_{\rm rad}\,\psi(\rho) \ = \ \big[\,-\De_{\rm rad}(\rho) \ + \ V(\rho)\, \big]\,\psi(\rho)\ = \ E\, \psi(\rho)\ ,
\end{equation}
where the radial operator defined by
\[
  \De_{\rm rad}(\rho)\ =\ 2\,\bigg[
  \frac{1}{\mu_{13}} \rho_{13}\, \pa_{\rho_{13}}^2 +
  \frac{1}{\mu_{23}} \rho_{23}\, \pa_{\rho_{23}}^2 +
  \frac{1}{\mu_{12}} \rho_{12}\,\pa_{\rho_{12}}^2 +
\]
\[
  \frac{(\rho_{13} + \rho_{12} - \rho_{23})}{m_1}\pa_{\rho_{13},\,\rho_{12}} +
  \frac{(\rho_{13} + \rho_{23} - \rho_{12})}{m_3}\pa_{\rho_{13},\,\rho_{23}} +
  \frac{(\rho_{23} + \rho_{12} - \rho_{13})}{m_2}\pa_{\rho_{23},\,\rho_{12}}\bigg] +
\]
\begin{equation}
\label{addition3-3r-M}
  \frac{d}{\mu_{13}} \pa_{\rho_{13}} +
  \frac{d}{\mu_{23}} \pa_{\rho_{23}} +
  \frac{d}{\mu_{12}} \pa_{\rho_{12}}\ ,
\end{equation}
governs the kinetic (radial) dynamics of the relative motion, and
\begin{equation}
\label{mus}
   \mu_{ij}\ =\ \frac{m_i\,m_j}{m_i+ m_j}\ ,
\end{equation}
is the reduced mass for particles $i$ and $j$. The radial Hamiltonian (\ref{Hrad3}) is equivalent, up to a gauge transformation, to a three-dimensional Schr\"odinger operator, see Ref. \cite{TME3-d}.

\emph{The operator (\ref{Hrad3}) describes all the eigenfunctions with zero angular momentum of the original Hamiltonian (\ref{Hgen})}. In particular, it describes the ground state level we usually are interested in.

For the operator ${\cal H}_{\rm rad}$ (\ref{Hrad3}), the configuration space is given by $S_{\Delta}\geq0$ where

\begin{equation}\label{}
  S_{\Delta} \ \equiv \ \frac{1}{4}\sqrt{2\,(\rho_{12}\,\rho_{13}\,+\,\rho_{12}\,\rho_{23}+\rho_{13}\,\rho_{23})\,-\,(\rho_{12}^2\,+\,\rho_{13}^2\,
+\,\rho_{23}^2) }\ ,
\end{equation}

is the area of the \emph{triangle of interaction} whose vertices are the individual positions $\mathbf{r}_i$ of the three particles.

The reduced Hamiltonian ${\cal H}_{\rm rad}$ (\ref{Hrad3}) is essentially self-adjoint with respect to the radial measure
\begin{equation}
\label{normmeasure}
d\varrho \ =\ {(S_{\Delta})}^{d-3}\,d\rho_{12}\,d\rho_{13}\,d\rho_{23}\ .
\end{equation}

Although the radial Hamiltonian ${\cal H}_{\rm rad}$ is self-adjoint it is not in the form of a Laplace-Beltrami operator plus a potential. For this to be true a further $d-$dependent gauge transformation is needed, as shown in Ref. \cite{TME3-d}. For $d=3$, a case partially studied in Ref. \cite{Moshinski}, the radial measure (\ref{normmeasure}) is greatly simplified.

\section{Three-body harmonic oscillator}

Now, let us consider the Hamiltonian (\ref{Hrad3}) with potential
\begin{equation}
\label{V3-es}
   V^{(es)}\ =\ 2\,\om^2\bigg[   \nu_{12}\,\rho_{12} \ + \  \nu_{13}\,\rho_{13} \ + \  \nu_{23}\,\rho_{23} \bigg]\ ,
\end{equation}
where $\om$ and $\nu_{12},\,\nu_{13},\,\nu_{23}$ are positive constants. Equivalently, in terms of the relative distances $r_{ij}$ (\ref{redi}) between particles, (\ref{V3-es}) is an harmonic pairwise potential.
It is easy to verify that the eigenfunctions $\Psi$ of the radial Hamiltonian (\ref{Hrad3}) with potential ($\ref{V3-es}$), 
\begin{equation}
\label{H3-es}
   {\cal H}^{(es)}_{\rm rad}\ = \ -\De_{\rm rad}(\rho)\ + \  2\,\om^2\bigg[   \nu_{12}\,\rho_{12} \ + \  \nu_{13}\,\rho_{13} \ + \  \nu_{23}\,\rho_{23} \bigg]\ ,
\end{equation}
occur in the form
\[
\Psi(\rho_{12},\,\rho_{13},\,\rho_{23}) \ = \ \Psi_0^{(es)}(\rho_{12},\,\rho_{13},\,\rho_{23}) \ \times P_N(\rho_{12},\,\rho_{13},\,\rho_{23}) \ ,
\]
where $\Psi_0^{(es)}$ (the ground state) is a global common factor and $P_N$ is a multivariable polynomial function in the $\rho$-variables \cite{TME3-d}.
Its spectra is linear in quantum numbers. Moreover, the operator (\ref{H3-es}) is exactly solvable which implies that one can compute the spectrum and the eigenfunctions by pure algebraic methods. In particular, the ground state can be taken in the following form
\begin{equation}
\label{Psi03}
   \Psi_0^{(es)}\ =\ {\cal N}\,e^{-\om\, (a\,\mu_{12}\,\rho_{12}\,+\,b\,\mu_{13}\,\rho_{13}\,+\,c\,\mu_{23}\,\rho_{23})}\ ,
\end{equation}
where ${\cal N}$ is a normalization factor. The parameters $a,\,b,\,c$ in the exponent can be related to those of the potential ($\ref{V3-es}$) through the algebraic equations
\begin{equation}
\label{freq-3}
\begin{aligned}
& \nu_{12} \ = \  a^2\, \mu _{12} \ + \ a\, b\ \frac{\mu _{12}\, \mu _{13} }{m_1} \ + \  a\, c\ \frac{\mu _{12} \,\mu _{23} }{m_2} \ - \  b\, c \ \frac{\mu_{13} \,\mu _{23} }{m_3} \ ,
\\ &
\nu_{13} \ = \  b^2\, \mu _{13} \ + \  a\, b\ \frac{\mu _{12} \,\mu _{13} }{m_1}\ + \  b \,c\ \frac{\mu _{13} \,\mu _{23}}{m_3}\  - \ a\, c\ \frac{\mu _{12} \,\mu _{23} }{m_2} \ ,
\\ &
\nu_{23} \ = \ c^2 \,\mu _{23} \ + \  a \,c \ \frac{\mu _{12} \,\mu _{23} }{m_2}\ + \  b\, c\ \frac{\mu _{13} \,\mu _{23} }{m_3} \ - \   a\, b\ \frac{\mu _{12}\, \mu _{13} }{m_1} \ ,
\end{aligned}
\end{equation}

such that the ground state energy is given by
\begin{equation}
\label{E00}
E_0^{(es)}\ = \ \om \,d\,(a+b+c) \ .
\end{equation}

In the three $\rho$-variables, the exactly-solvable Hamiltonian (\ref{H3-es}) does not admit separation of variables. However, the ground state eigenfunction (\ref{Psi03}) can be trivially factored as the product of three functions, each of them depending on a single $\rho$ variable.

\emph{By construction, the eigenfunctions of the three-dimensional Hamiltonian (\ref{H3-es}) are also eigenfunctions (with the same energy) of the original $3d$-dimensional Hamiltonian (\ref{Hgen}).}

\subsection{Case of equal masses}

In this Section we consider the case of three particles of equal masses $m_1=m_2=m_3=m$, but arbitrary constants $a,b,c>0$. The harmonic oscillator potential (\ref{V3-es}) becomes
\begin{equation}
\label{V3equal}
   V^{(3m)}\ =\ \frac{1}{2}\,m\,\om^2\,[\,(2 \,a^2+a (b+c)-b\, c)\,\rho_{12} \ + \ (2\,b^2+b(a+c)-a\, c)\,\rho_{13} \ + \ (2\,c^2+c(a+b)-a\, b)\,\rho_{23}\,] \ .
\end{equation}
It is a type of non-isotropic 3-body harmonic oscillator with different spring constants.
In this case the exact ground state function (\ref{Psi03}) and the associated energy (\ref{E00}) are given by
\begin{equation}
\label{Psi03m}
   \Psi_0^{(3m)}\ =\ e^{-\frac{\om\,m }{2}\,(\,a\,\rho_{12}\ +\ b\,\rho_{13}\ +\ c\,\rho_{23}\,)}\ ,
\end{equation}
\begin{equation}
\label{E03m}
E_0^{(3m)}\ = \ \om \,d\,(a+b+c) \ .
\end{equation}

The case of 3 identical spring constants in (\ref{V3equal}) corresponds to $a=b=c$.

\subsection{Case of two equal massive particles}

Now, we move to the case where two of the three particles are identical, i.e. we put
\[
m_1\ = \ m_2 \ = \ 1 \ ; \qquad  m_3 \ = \ m \ ,
\]
and interact through an harmonic oscillator potential, namely

\begin{equation}
\label{V3-esBO}
   V\ = \    \frac{1}{4}\,\rho_{12} \ + \  \frac{1}{2}\,K\,\rho_{13} \ + \  \frac{1}{2}\,K\,\rho_{23}\ , \qquad K > 0 \ .
\end{equation}

By putting
\[
a\ = \ \frac{1}{2}\left(\sqrt{K+1}\,-\,\sqrt{\frac{K\,m}{m+2}}\right)\qquad , \quad b=c=\frac{\sqrt{K} (m+1)}{2  \sqrt{m(m+2)}}\qquad , \quad  \omega=1 \ ,
\]
in Eqs. (\ref{V3-es}) and (\ref{freq-3})  we arrive to the expression (\ref{V3-esBO}). For the three-dimensional case $d=3$, this physically important problem was studied in Ref. \cite{Moshinski}. In order to make a comparison, we adopted the same units of mass and spring constants used in Ref. \cite{Moshinski}.

\subsubsection{Exact result}

The exact ground state energy (\ref{E00}) and the eigenfunction (\ref{Psi03}) reduce to

\begin{equation}\label{E0}
E_0\ = \ \frac{1}{2}\, d\, \left(\sqrt{\frac{K\,(m+2)}{m}}+\sqrt{K+1}\right) \ ,
\end{equation}

\begin{equation}\label{P0}
 \psi_0 \ = \ {\bigg(\frac{\sqrt{\pi}\, \Gamma \big(\frac{d}{2}\big)\, \Gamma \big(\frac{d-1}{2}\big)}{2^{d-4}}\bigg)}^{-\frac{1}{2}}\,{\bigg(\frac{K\,m}{m+2}\,(1+K)\bigg)}^{\frac{d}{8}} e^{ \frac{1}{4} \left( \sqrt{\frac{K\,m}{m+2}} \left(\rho _{12}-2 \left(\rho _{13}+\rho _{23}\right)\right)-\sqrt{K+1} \rho _{12}\right)} \ ,
\end{equation}

respectively. The function (\ref{P0}) is normalized with respect to the radial measure $d\varrho$ (\ref{normmeasure}).

The following remark is in order. For the three-dimensional case $d=3$, Moshinsky and Kittel \cite{Moshinski} studied the original Hamiltonian (\ref{Hgen}) with potential (\ref{V3-esBO}). After separation of the center of mass, the 9-dimensional problem reduces to a 6-dimensional one in the space of relative motion. In this space, they introduce two 3-dimensional vectorial Jacobi coordinates ${\bf r}_1^{(J)}$ and ${\bf r}_2^{(J)}$. Hence, for the normalization of the eigenfunctions they do not use (\ref{normmeasure}) but the factorizable integration measure
\begin{equation}\label{MeMK}
d^3{\bf r}_1^{(J)}\,d^3{\bf r}_2^{(J)} \ = \ d\Omega\,d\varrho \ ,
\end{equation}
where $d\Omega$ involves 3 angular variables alone.

Now, the ground state function of (\ref{Hgen}) must depend on the relative distances only \cite{Ter}. This fundamental fact is not evident in \cite{Moshinski}, whereas in the present formalism it does, see eq. (\ref{P0}). Using the measure (\ref{MeMK}), immediately we see that the ground state eigenfunction (\ref{P0}) reproduces up to the corresponding constant factor coming from the trivial integration over $d\Omega$, the result reported in \cite{Moshinski}. The energy (\ref{E0}) is exactly the same value obtained in \cite{Moshinski}, as it should be. That way, we nicely reproduce the results presented in \cite{Moshinski} and extend them to arbitrary dimension $d$.

\subsubsection{Approximate solution}

As for the Born-Oppenheimer approximation one starts with the assumption that the two identical particles $m_1=m_2=1$ are much heavier than the third one $m$, thus $m\ll 1$. The masses $m_1$ and $m_2$ are fixed at positions $\mathbf{r}_1$, $\mathbf{r}_2$, respectively. In this case $\rho_{12}$ is also fixed, and one solves first the electronic radial Hamiltonian

\begin{equation}
\label{H-3-bodyelec}
   {\cal H}^{{\rm (electronic)}}_{\rm rad} \psi^{(e)}\  \equiv \  \big[\,-\De^{{\rm (electronic)}}_{\rm rad}\ + \  \frac{1}{2}\,K\,\rho_{13} \ + \  \frac{1}{2}\,K\,\rho_{23} \,\big]\,\psi^{(e)}\ =\ E^{(e)}\, \psi^{(e)}\ ,
\end{equation}
where the operator
\begin{equation}
\label{addition3-3r-Melec}
 \De^{{\rm (electronic)}}_{\rm rad}\ =\ \frac{2}{m}\,\bigg[
  \rho_{13}\, \pa_{\rho_{13}}^2 +
   \rho_{23}\, \pa_{\rho_{23}}^2 +
  (\rho_{13} + \rho_{23} - \rho_{12})\,\pa_{\rho_{13},\,\rho_{23}} +   \frac{d}{2} \pa_{\rho_{13}} +
  \frac{d}{2} \pa_{\rho_{23}} \bigg] \ .
\end{equation}

depends on $\rho_{12}$ parametrically, in other words $\rho_{12}$ plays the role of a classical variable. Formally, the operator (\ref{addition3-3r-Melec}) can be obtained from ({\ref{addition3-3r-M}) in the limit $m_1=m_2\rightarrow \infty$ together with $m_3\rightarrow m$.

For the electronic eigenvalue problem (\ref{H-3-bodyelec}) we obtain the ground state function

\begin{equation}\label{pelect}
\psi^{(e)}_0 \  = \ {\pi}^{-\frac{d}{4}}\, {(2\,K\,m)}^{\frac{d}{8}}  \,  e^{\frac{1}{4}\sqrt{\frac{K\,m}{2}}(\rho_{12} \,-\,2(\rho_{13}\,+\,\rho_{23}))}  \ ,
\end{equation}

with energy

\begin{equation}\label{Eelec}
E^{(e)}_0 \  = \ d\,\sqrt{\frac{K}{2\,m}}\ + \ \frac{K \,\rho_{12}}{4} \ .
\end{equation}

The ground state (\ref{pelect}) obeys the $L^2$-condition

\begin{equation}\label{IC}
\int {\big(\psi^{(e)}_0\big)}^2\,d\mathbf{r}_{{}_3}\ = \ 1 \ ,
\end{equation}

where the integration is over the electronic variable $\mathbf{r}_{{}_3}\in \mathbb{R}^d$ only.

For $d=3$, using standard spherical coordinates $\mathbf{r}_{{}_3}=({r}_{{}_3},\,\theta,\,\phi)$ we easily find the relation  $d^3\mathbf{r}_{{}_3}={r}_{{}_3}^2\,\sin\theta\, d{r}_{{}_3}\,d\theta\, d\phi=\frac{1}{4\,\sqrt{\rho_{12}}}\,d\rho_{13}\,d\rho_{23}\,d\phi$. Then, at $d=3$ the solution (\ref{pelect}) coincides with that presented in Ref. \cite{Moshinski}.

Expression (\ref{Eelec}), under the name of a \emph{potential curve}, is the one we must add to the Hamiltonian of the two heavy particles $m_1=m_2=1$, namely $(-4\,\rho_{12}\,\pa_{\rho_{12}}^2 -
 2\,d\, \pa_{\rho_{12}} + \frac{1}{4}\,\rho_{12})$, to get the nuclear Hamiltonian, i.e.

\begin{equation}\label{Hnuclear}
  {\cal H}^{\rm (nuclear)}\ =\ \bigg[-4\,\rho_{12}\,\pa_{\rho_{12}}^2-
 2\,d\, \pa_{\rho_{12}} + \frac{1}{4}\,\rho_{12} \bigg]\ + \  d\,\sqrt{\frac{K}{2\,m}} \  + \   \frac{K \,\rho_{12}}{4}\ .
\end{equation}

Clearly, the ground state of (\ref{Hnuclear})
\[
\psi^{(n)}_0 \ \propto \ e^{-\frac{\sqrt{ (K+1)}}{4 }\,\rho_{12}} \ ,
\]

is the one of zero quanta with frequency $\sqrt{K+1}$. Therefore, the normalized total wave function in the Born-Oppenheimer approximation (BOA) takes the form

\begin{equation}\label{PBOP}
\psi^{\rm (BO)}_0 \  = \   {\bigg(\frac{\sqrt{\pi}\, \Gamma \big(\frac{d}{2}\big)\, \Gamma \big(\frac{d-1}{2}\big)}{2^{d-4}}\bigg)}^{-\frac{1}{2}}\,{\bigg(\frac{K\,(1+K)\,m}{2}\bigg)}^{\frac{d}{8}} \,e^{-\frac{\sqrt{ (K+1)} }{4 }\,\rho_{12}}\,\times\,  e^{\frac{1}{4}\sqrt{\frac{K\,m}{2}}(\rho_{12} \,-\,2(\rho_{13}\,+\,\rho_{23}))}
\end{equation}
\[
\qquad \ \propto \ \psi^{(e)}_0 \times \psi^{(n)}_0 \ ,
\]

with respect to the measure (\ref{normmeasure}). The corresponding ground state energy is given by

\begin{equation}\label{E0B03}
E^{\rm (BO)}_0  \ = \   \frac{1}{2} \,d\, \left(\sqrt{\frac{{2\,K}}{m}}+\sqrt{K+1}\right) \ .
\end{equation}

This expression corresponds to the generalization to $d$-dimensions of the result obtained in Ref. \cite{Moshinski}.

\subsubsection{Accuracy of the Born-Oppenheimer approximation}

Assuming $m\ll1$, we first compare the energies $E_0$ (\ref{E0}) and $E^{\rm (BO)}_0$ (\ref{E0B03})

\begin{equation}\label{delE}
\begin{aligned}
\Delta E \equiv \ \frac{E_0 \ - \ E^{\rm (BO)}_0}{E_0}  \ & = \ 1 \ - \ \frac{E^{\rm (BO)}_0}{E_0}  \ = \ 1 \ - \ \frac{ \sqrt{\frac{{2\,K}}{m}}+\sqrt{K+1}}{ \sqrt{\frac{K\,(m+2)}{m}}+\sqrt{K+1}}
\\ &
\approx \   \ \frac{1}{4}\,m \ - \   \frac{1}{4} \sqrt{\frac{K+1}{2\,K}}\, m^{3/2}\ + \ \frac{(K+4)}{32\, K}\, m^2\ + \ \dots  \ .
\end{aligned}
\end{equation}

\begin{figure}[h]
  \centering
  \includegraphics[width=7.0cm]{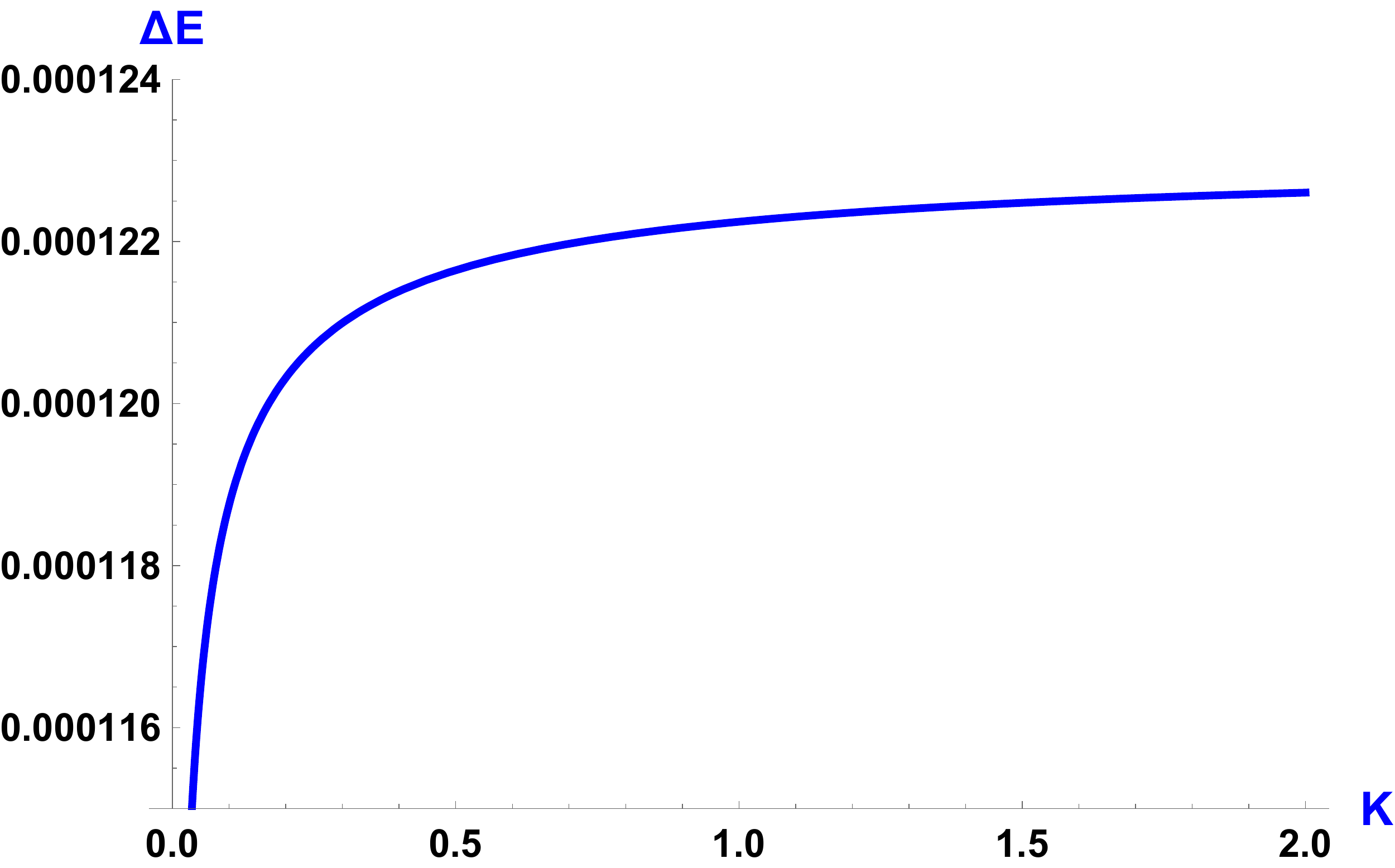} \qquad  \includegraphics[width=7.0cm]{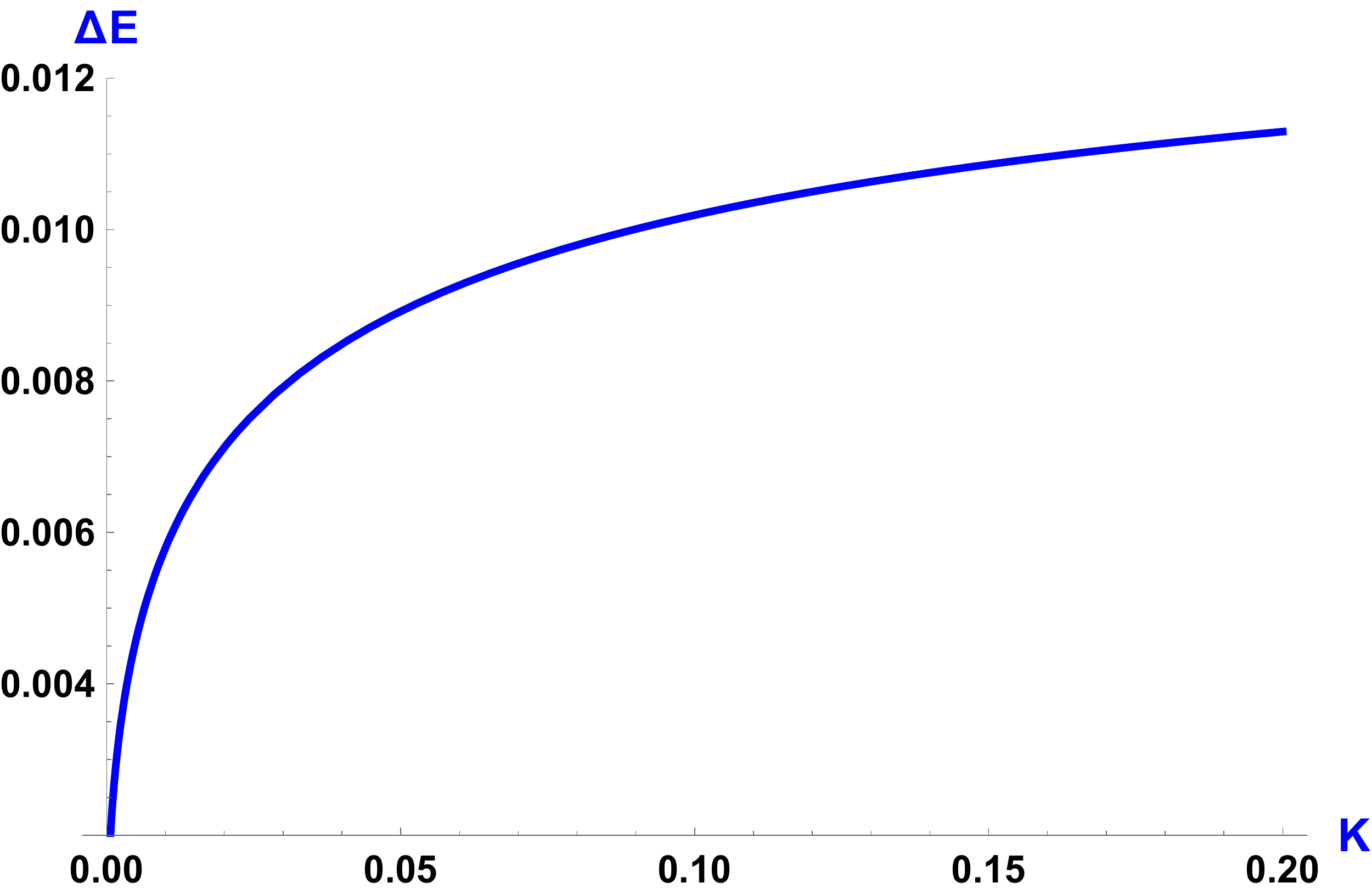}
  \caption{The ratio $\Delta E$ (\ref{delE}) as a function of the spring constant $K$ for (left) the hydrogen molecular ion $H_2^+$ ($m=\frac{1}{2000}$) and (right) the proton in a hydrogen bond between two nitrogen or oxygen atoms ($m=\frac{1}{15}$). The ratio $\Delta E$ is a monotonic increasing (bounded) function of $K$. }\label{G1}
\end{figure}

Remarkably, the ratio $\Delta E$ (\ref{delE}) does not depend on the dimension $d$. In Figure \ref{G1} we plot $\Delta E$ (\ref{delE}) as a function of the spring constant $K$ in the case of two relevant systems, namely the hydrogen molecular ion $H_2^+$ ($m=\frac{1}{2000}$) and the proton in a hydrogen bond between two nitrogen or oxygen atoms ($m=\frac{1}{15}$).

In powers of the small parameter $m\ll1$ we have

\begin{equation}\label{E0B0ex}
E_0  \ = \   E^{\rm (BO)}_0 \ + \ \frac{d\,\sqrt{K}}{4\,\sqrt{2}}\,\bigg( \,m^{\frac{1}{2}}\ - \   \frac{1}{8}\,m^{\frac{3}{2}} \ + \   \frac{1}{32}\,m^{\frac{5}{2}} \ + \ \dots \bigg)  \ ,
\end{equation}

hence, $E^{\rm (BO)}_0$ is nothing but the sum of the first two terms of the Puiseux series of the exact result $E_0$ (\ref{E0}). Similarly, we obtain that the exponent in (\ref{PBOP}), i.e. the phase
\[
\Phi_{\rm BO}\ \equiv \ {\frac{1}{4}\sqrt{\frac{K\,m}{2}}(\rho_{12} \,-\,2(\rho_{13}\,+\,\rho_{23}))}-\frac{\sqrt{ (K+1)} }{4 }\,\rho_{12}
\]
of the ground state function in the Born-Oppenheimer approximation, coincides exactly with the sum of the first two terms of the Puiseux series of the exact result (\ref{P0})
\[
\Phi \ =\ \frac{1}{4} \left( \sqrt{\frac{K\,m}{m+2}} \left(\rho _{12}-2 \left(\rho _{13}+\rho _{23}\right)\right)-\sqrt{K+1} \rho _{12}\right) \ .
\]
Explicitly,
\begin{equation}\label{}
\Phi  \ = \   \Phi_{\rm BO} \ - \ \frac{1}{16}\sqrt{\frac{K}{2}} \left(\rho _{12}-2 \left(\rho _{13}+\rho _{23}\right)\right)\bigg(\,m^{\frac{3}{2}} \ -\ \frac{3}{8}\,m^{\frac{5}{2}} \ + \ {\cal O}(\,m^{\frac{7}{2}}) \bigg) \ .
\end{equation}

To compare the wave functions, we consider the overlap $T\equiv {\langle\psi^{\rm (BO)}_0\vert \psi_0\rangle}^2$, which from
(\ref{PBOP}) and (\ref{P0}) is given by

\begin{equation}\label{overlap}
T = \   2^{\frac{7\, d}{4}}\, (m+2)^{d/4}\, \left(\sqrt{2\,(m+2)}+2\right)^{-d} \  \simeq  \ 1 \ - \ d\frac{ m^2}{128} \ + \ d\frac{m^3}{256} \ + \  {\cal O}(m^4)  \ .
\end{equation}

The overlap $T$ does not depend on the spring constant $K$. However, it depends on the dimension $d$.

\begin{figure}[h]
  \centering
  \includegraphics[width=13.0cm]{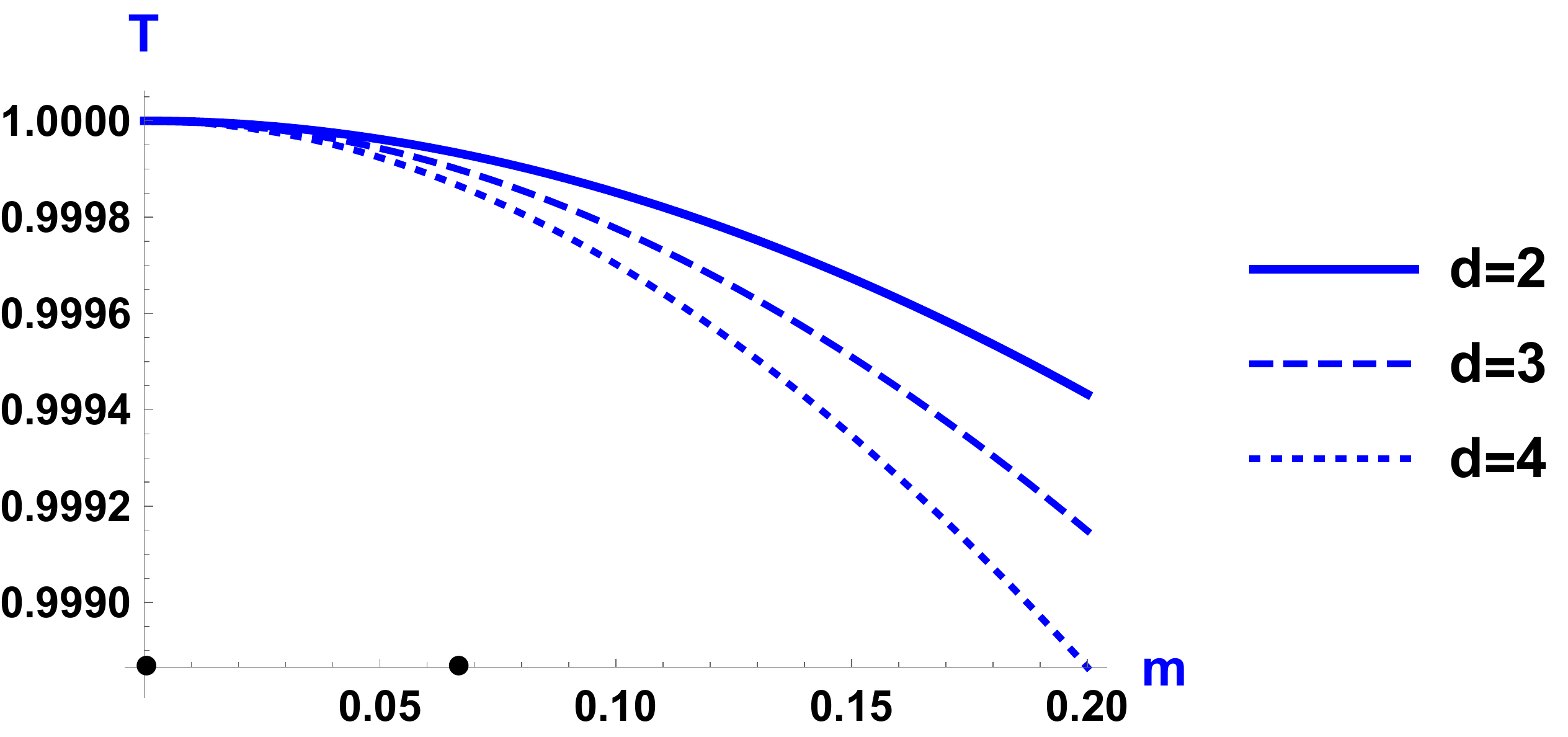}
  \caption{The overlap $T$ (\ref{overlap}) as a function of the mass $m$ of the light particle at $d=2,3,4$. The two marked circles stand for the hydrogen molecular ion $H_2^+$ ($m=\frac{1}{2000}$) and for the proton in a hydrogen bond between two nitrogen or oxygen atoms ($m=\frac{1}{15}$), respectively.}\label{G2}
\end{figure}

For the physically important cases of $H_2^+$ ($m=\frac{1}{2000}$) and hydrogen bond ($m=\frac{1}{15}$) in three-dimensions $d=3$, the overlap (\ref{overlap}) differs from $1$ by terms of the order less than one part in $10^{8}$ and four parts in $10^4$, respectively \cite{Moshinski}. The new results (\ref{E0B0ex})-(\ref{overlap}) show that also for the physically relevant two-dimensional case $d=2$, the Born-Oppenheimer approximation provides accurate values for the energy as well as for the ground state wave function.

\section{Four body case }

Similarly to the three-body case, assuming $d\geq 3$ and that the potential $V=V(\rho)$ solely depends on the six independent relative variables
\[
    \rho_{12}=r_{12}^2\ ,\quad  \rho_{13}=r_{13}^2\ ,\quad \rho_{23}=r_{23}^2\ ,\quad \rho_{14}=r_{14}^2\ ,\quad \rho_{24}=r_{24}^2\ ,\quad \rho_{34}=r_{34}^2 \ ,
\]

one arrives at the six-dimensional radial equation in the $\rho$-space of the relative motion
\begin{equation}
\label{H-4-body}
       \,[-\De^{(4)}_{\rm rad}(\rho) \ + \ V(\rho) \,]\,\psi(\rho)\ =\ E\, \psi(\rho)\ ,
\end{equation}
where

{\small
\begin{equation}
\begin{aligned}
\label{addition4-3r-M}
& { \De^{(4)}_{\rm rad}}(\rho)\ =  \   2\,\bigg[ \frac{1}{\mu_{12}} \rho_{12} \,\pa^2_{\rho_{12}}\ + \ \frac{1}{\mu_{13}}\rho_{13}\, \pa^2_{\rho_{13}}\  + \ \frac{1}{\mu_{14}}\rho_{14}\, \pa^2_{\rho_{14}}
 + \ \frac{1}{\mu_{23}}\rho_{23}\, \pa^2_{\rho_{23}}
\\ &  + \ \frac{1}{\mu_{24}}\rho_{24}\, \pa^2_{\rho_{24}}
 \ + \ \frac{1}{\mu_{34}}\rho_{34}\, \pa^2_{\rho_{34}} \bigg]
\\ &
 +  \frac{2}{m_1} \bigg({(\rho_{12} + \rho_{13} - \rho_{23})}\pa_{\rho_{12}}\pa_{\rho_{13}}\ +
          {(\rho_{12} + \rho_{14} - \rho_{24})}\pa_{\rho_{12}}\pa_{\rho_{14}}\ +
          {(\rho_{13} + \rho_{14} - \rho_{34})}\pa_{\rho_{13}}\pa_{\rho_{14}} \bigg)
\\ &
+  \frac{2}{m_2} \bigg((\rho_{12} + \rho_{23} - \rho_{13})\pa_{\rho_{12}}\pa_{\rho_{23}}\ +
          (\rho_{12} + \rho_{24} - \rho_{14})\pa_{\rho_{12}}\pa_{\rho_{24}}\ +
          (\rho_{23} + \rho_{24} - \rho_{34})\pa_{\rho_{23}}\pa_{\rho_{24}}
    \bigg)
\\ &
+  \frac{2}{m_3} \bigg((\rho_{13} + \rho_{23} - \rho_{12})\pa_{\rho_{13}}\pa_{\rho_{23}}\ +
          (\rho_{13} + \rho_{34} - \rho_{14})\pa_{\rho_{13}}\pa_{\rho_{34}}\ +
          (\rho_{23} + \rho_{34} - \rho_{24})\pa_{\rho_{23}}\pa_{\rho_{34}}
    \bigg)
\\ &
+  \frac{2}{m_4} \bigg((\rho_{14} + \rho_{24} - \rho_{12})\pa_{\rho_{14}}\pa_{\rho_{24}}\ +
          (\rho_{14} + \rho_{34} - \rho_{13})\pa_{\rho_{14}}\pa_{\rho_{34}}\ +
          (\rho_{24} + \rho_{34} - \rho_{23})\pa_{\rho_{24}}\pa_{\rho_{34}}
    \bigg)
\\ &
\ + \ d\,\bigg[\frac{1}{\mu_{12}}\pa_{\rho_{12}} + \frac{1}{\mu_{13}}\pa_{\rho_{13}}+ \frac{1}{\mu_{14}}\pa_{\rho_{14}}+ \frac{1}{\mu_{23}}\pa_{\rho_{23}}+ \frac{1}{\mu_{24}}\pa_{\rho_{24}}+ \frac{1}{\mu_{34}}\pa_{\rho_{34}}\bigg]         \ ,
\end{aligned}
\end{equation}
}
 where $\mu_{ij}$ is defined in Eq. (\ref{mus}) and $\De^{(4)}_{\rm rad}$ plays the role of kinetic radial operator cf.(\ref{addition3-3r-M}). The operator
\begin{equation}
\label{Hrad4}
  {\cal H}^{(4)}_{\rm rad}\ \equiv\ -\De^{(4)}_{\rm rad} \ + \  V\ ,
\end{equation}
is equivalent to a six-dimensional radial Schr\"odinger operator, for further details see \cite{TME4-d}. It can be called six-dimensional radial Hamiltonian. As a function of the six $\rho$-variables, the operator (\ref{addition4-3r-M}) is not $S_6$ permutationally-invariant. Nevertheless, it remains $S_4$ invariant under the permutations of the particles. For the three-body case, where the number of $\rho$ variables (relative distances squared) equals the number of particles, the corresponding operator $\Delta_{\rm rad}$ is indeed $S_3$ permutationally-invariant.

For the Hamiltonian ${\cal H}^{(4)}_{\rm rad}$ (\ref{Hrad4}), the configuration space is given by ${\cal V}\geq0$ where

\begin{equation}
\label{Vol4}
\begin{aligned}
 {\cal V}  \ \equiv \ & \frac{1}{12} \bigg[\,\left[\left(\rho _{13}+\rho _{14}+\rho _{23}+
  \rho _{24}\right) \rho _{34}-\left(\rho _{13}-\rho _{14}\right) \left(\rho _{23}-\rho _{24}\right)-
  \rho _{34}^2 \right] \rho _{12}
\\ & - \ \rho _{13}^2 \rho _{24} \ - \ \rho _{34} \rho _{12}^2 \ +
\  \rho _{23} \left[\left(\rho _{14}-\rho _{24}\right) \rho _{34}-\rho _{14} \left(\rho _{14}+
  \rho _{23}-\rho _{24}\right)\right]
\\ &  \ + \ \rho _{13} \left[\,\rho _{14} \left(\rho _{23}+\rho _{24}-
  \rho _{34}\right)+\rho _{24} \left(\rho _{23}-\rho _{24}+\rho _{34}\right)\right]\bigg]^{\frac{1}{2}}  \ ,
\end{aligned}
\end{equation}
is the volume of the \emph{tetrahedron of interaction} whose vertices correspond to the positions of the particles.

The reduced Hamiltonian ${\cal H}^{(4)}_{\rm rad}$ (\ref{Hrad4}) is essentially self-adjoint with respect to the radial measure
\begin{equation}
\label{normmeasure4}
d\varrho \ =\ {\cal V}^{ d-4}\,d\rho_{12}\,d\rho_{13}\,d\rho_{14}\,d\rho_{23}\,d\rho_{24}\,d\rho_{34}\ .
\end{equation}

Although the radial Hamiltonian is essentially self-adjoint it is not in the form of a Laplace-Beltrami operator plus potential. For this to be true a further gauge transformation is needed, see \cite{TME4-d}.

\section{Four-body harmonic oscillator}

For the four-body system, let us introduce the harmonic potential
\begin{equation}
\label{V4-es}
   \tilde V^{(es)}\ =\ 2\,\om^2\bigg[   \nu_{12}\,\rho_{12} \ + \  \nu_{13}\,\rho_{13} \ + \  \nu_{14}\,\rho_{14} \ + \  \nu_{23}\,\rho_{23} \ + \  \nu_{24}\,\rho_{24} \ + \  \nu_{34}\,\rho_{34}  \bigg]\ ,
\end{equation}
where $\om$ and the $\nu$'s are positive constants. It is easy to verify that the eigenfunctions of the Hamiltonian (\ref{Hrad4}) with potential ($\ref{V4-es}$)
\begin{equation}
\label{H4es}
   \tilde {\cal H}^{(es)}_{\rm rad} \ = \ - \De^{(4)}_{\rm rad}(\rho)\ + \ \tilde V^{(es)}  \ ,
\end{equation}
occur in the form
\[
\tilde \Psi(\rho) \ = \ \tilde \Psi_0^{(es)}(\rho) \ \times \tilde P_N(\rho) \ ,
\]
where $\tilde \Psi_0^{(es)}$ (the ground state) is a global common factor and $\tilde P_N$ is a multivariable polynomial function in the six $\rho$-variables.
Its spectra is linear in quantum numbers. Again, the operator (\ref{H4es}) is exactly solvable. The ground state function takes the following form
\begin{equation}
\label{Psi04}
   \tilde \Psi_0^{(es)}\ =\ {\cal N}\,e^{-\om\, (a\,\mu_{12}\,\rho_{12}\,+\,b\,\mu_{13}\,\rho_{13}\,+\,c\,\mu_{14}\,\rho_{14}\,+\,e\,\mu_{23}\,\rho_{23}\,+\,f\,\mu_{24}\,\rho_{24}
   \,+\,g\,\mu_{34}\,\rho_{34})}\ ,
\end{equation}
where ${\cal N}$ is a normalization factor and the parameters $a,\,b,\,c,\,e,\,f,\,g$ in the exponent are related to those of the potential ($\ref{V4-es}$) through the six algebraic equations
\begin{equation}
\label{freq-4}
\begin{aligned}
& \nu_{12} \ = \  a^2\, \mu _{12} \ + \ a\, b\ \frac{\mu _{12}\, \mu _{13} }{m_1} \ + \  a\, c\ \frac{\mu _{12} \,\mu _{14} }{m_1}\ + \  a\, e\ \frac{\mu _{12} \,\mu _{23} }{m_2}\ + \  a\, f\ \frac{\mu _{12} \,\mu _{24} }{m_2}
\\ &
\ - \  b\, e \ \frac{\mu_{13} \,\mu _{23} }{m_3}\ - \  c\, f \ \frac{\mu_{14} \,\mu _{24} }{m_4} \ ,
\\ &
\nu_{13} \ = \  b^2\, \mu _{13} \ + \ b\, a\ \frac{\mu _{13}\, \mu _{12} }{m_1} \ + \  b\, c\ \frac{\mu _{13} \,\mu _{14} }{m_1}\ + \  b\, e\ \frac{\mu _{13} \,\mu _{23} }{m_3}\ + \  b\, g\ \frac{\mu _{13} \,\mu _{34} }{m_3}
\\ &
\ - \  a\, e \ \frac{\mu_{12} \,\mu _{23} }{m_2}\ - \  c\, g \ \frac{\mu_{14} \,\mu _{34} }{m_4} \ ,
\\ &
\vdots
\\ &
\nu_{34} \ = \  g^2\, \mu _{34} \ + \ g\, b\ \frac{\mu _{34}\, \mu _{13} }{m_3} \ + \  g\, c\ \frac{\mu _{34} \,\mu _{14} }{m_4}\ + \  g\, e\ \frac{\mu _{34} \,\mu _{23} }{m_3}\ + \  g\, f\ \frac{\mu _{34} \,\mu _{24} }{m_4}
\\ &
\ - \  b\, c \ \frac{\mu_{13} \,\mu _{14} }{m_1}\ - \  e\, f \ \frac{\mu_{23} \,\mu _{24} }{m_4}
\ .
\end{aligned}
\end{equation}
and the ground state energy takes the simple form
\begin{equation}
\label{E004}
\tilde E_0^{(es)}\ = \ \om \,d\,(a+b+c+e+f+g) \ .
\end{equation}

\subsection{Case of equal masses}

Let us consider the case of four particles of equal masses $m_1=m_2=m_3=m_4=m$, but arbitrary constants $a,b,c,e,f,g>0$. From (\ref{freq-4}), it follows that the harmonic oscillator potential (\ref{V4-es}) reduces to
\begin{equation}
\label{V4equal}
\begin{aligned}
   V^{(4m)}\ & =\ \frac{1}{2}\,m\,\om^2\,\big[\ (2 a^2+a (b+c+e+f)-b e-c f)\,\rho_{12} \ + \ (2 b^2+b (a+c+e+g)-a e-c g)\,\rho_{13}
\\ &
\ + \ (2 c^2+c (a+b+f+g)-a f-b g)\,\rho_{14}   \ + \ (2 e^2+e (a+b+f+g)-a b-f g)\,\rho_{23}
   \\ &
    \ + \ (2 f^2+f (a+c+e+g)-a c-e g)\,\rho_{24}  \ + \ (2 g^2+g (b+c+e+f)-b c-e f)\,\rho_{24}  \ \big]\ .
\end{aligned}
\end{equation}
It is a type of non-isotropic 4-body harmonic oscillator with different spring constants.
In this case the exact ground state function (\ref{Psi04}) is given by
\begin{equation}
\label{Psi04m}
   \Psi_0^{(4m)}\ =\ e^{-\frac{\om\,m }{2}\,(\,a\,\rho_{12}\ +\ b\,\rho_{13}\ +\ c\,\rho_{14}\ +\ e\,\rho_{23}\ +\ f\,\rho_{24}\ +\ g\,\rho_{34}\,)}\ ,
\end{equation}
with energy
\begin{equation}
\label{E04m}
E_0^{(4m)}\ = \ \om \,d\,(a+b+c+e+f+g) \ .
\end{equation}

\subsection{Case of two equal massive particles}

\subsubsection{Exact result}

In this case we consider $d\geq 3$ and focus on the physically important case of two particles of equal mass ($m_1=m_2=1$) interacting between themselves and with another two particles ($m_3=m_4=m $) through an harmonic oscillator potential, namely

\begin{equation}
\label{V4}
   {\tilde V} \ =\    \frac{1}{4}\,\rho_{12} \ + \  \frac{K_1}{2}\,\rho_{34} \ + \  \frac{K_2}{2}(\,\rho_{13}  \ + \ \rho_{14} \ + \ \rho_{23} \ + \ \rho_{24}\,) \ , \qquad K_2 > 0 \, ; \, K_1 >0 \ .
\end{equation}

For the Hamiltonian (\ref{H4es}), the exact ground state energy and the corresponding eigenfunction are given by

\begin{equation}\label{E04}
\tilde E_0 \ = \ d\,(\,\alpha\, +\, 4\,\beta \, +\, {\gamma}) \ = \  \frac{1}{2}\, d\, \left( \sqrt{1+2\,K_2} \ +\ \sqrt{\frac{2\,(K_1+K_2)}{m}} \ +\ \sqrt{\frac{2\,K_2\,(1+m)}{m}} \right) \ ,
\end{equation}

\begin{equation}\label{P04}
\tilde \psi_0 \ = \ {\cal N}\, e^{\,-(\alpha\,\frac{1}{2}\,\rho_{12} \ + \ \beta\,\frac{ m}{(m+1)}\,[\rho_{13} \, + \,\rho_{14} \, + \,\rho_{23} \, + \,\rho_{24}] \ + \ \gamma\,\frac{m}{2}\,\,\rho_{34} \,)  } \ ,
\end{equation}

respectively, where

\begin{equation}\label{}
\begin{aligned}
& \alpha \ = \ \frac{1}{2} \bigg(  \sqrt{1\, + \, 2\,K_2} \ - \ \sqrt{\frac{2\,K_2\,m}{1+m}}   \bigg) \ ,
\\ &
 \beta \ = \ \frac{1}{2}\sqrt{\frac{K_2\,(1+m)}{2\,m}} \ ,
 \\ &
 \gamma \ = \ \frac{1}{\sqrt{2\,m}}\bigg(  \sqrt{K_1\, +\,K_2} \ - \ \sqrt{\frac{K_2}{1+m}}  \bigg) \ ,
\end{aligned}
\end{equation}

and ${\cal N}$ is a normalization constant.

\subsubsection{Approximate solution}

Now, for our problem at hand the Born-Oppenheimer approximation starts with the assumption that two masses $m_1=m_2=1$ are much heavier than the other two $m_3=m_4\ll 1$, thus $\mathbf{r}_1$ and $\mathbf{r}_2$ are fixed ($\rho_{12}$ is constant) and then one solves first the electronic radial  eigenvalue equation

\begin{equation}
\label{Hredelect4}
  \tilde {\cal H}^{{\rm (electronic)}}_{\rm rad}\,\tilde \psi^{(e)} \, \equiv \,    \big[\,-\tilde \De^{{\rm (electronic)}}_{\rm rad}\ + \   \frac{K_1}{2}\,\rho_{34} \ + \  \frac{K_2}{2}(\,\rho_{13}  \ + \ \rho_{14} \ + \ \rho_{23} \ + \ \rho_{24}\,) \,\big]\,\tilde \psi^{(e)}\ =\ \tilde E^{(e)}\, \tilde \psi^{(e)}\ ,
\end{equation}
where the operator

\begin{equation}
\begin{aligned}
\label{elect4}
& \frac{m}{2}\,\tilde \De^{{\rm (electronic)}}_{\rm rad}(\rho) \ =  \    \bigg[ \rho_{13}\, \pa^2_{\rho_{13}}\  + \ \rho_{14}\, \pa^2_{\rho_{14}}\ + \ \rho_{23}\, \pa^2_{\rho_{23}}\  + \ \rho_{24}\, \pa^2_{\rho_{24}}
 \ + \ 2\,\rho_{34}\, \pa^2_{\rho_{34}} \bigg]
\\ &
+ \bigg((\rho_{13} + \rho_{23} - \rho_{12})\pa_{\rho_{13}}\pa_{\rho_{23}}\ +
          (\rho_{13} + \rho_{34} - \rho_{14})\pa_{\rho_{13}}\pa_{\rho_{34}}\ +
          (\rho_{23} + \rho_{34} - \rho_{24})\pa_{\rho_{23}}\pa_{\rho_{34}}
    \bigg)
\\ &
+  \bigg((\rho_{14} + \rho_{24} - \rho_{12})\pa_{\rho_{14}}\pa_{\rho_{24}}\ +
          (\rho_{14} + \rho_{34} - \rho_{13})\pa_{\rho_{14}}\pa_{\rho_{34}}\ +
          (\rho_{24} + \rho_{34} - \rho_{23})\pa_{\rho_{24}}\pa_{\rho_{34}}
    \bigg)
\\ &
\ + \ \frac{d}{2}\,\bigg[ \pa_{\rho_{13}}+\pa_{\rho_{14}}+ \pa_{\rho_{23}}+\pa_{\rho_{24}}+ 2\,\pa_{\rho_{34}}\bigg]        \ ,
\end{aligned}
\end{equation}

depends on $\rho_{12}$ parametrically.
For the eigenvalue problem (\ref{Hredelect4}) we obtain the ground state function

\begin{equation}\label{Pe4}
\tilde \psi^{(e)}_0 \  = \ {\cal N}^{(e)}   \ e^{-\frac{1}{2} \sqrt{\frac{m}{2}}\big[\,\sqrt{K_2}\,(\rho_{13}+\rho_{14}+\rho_{23}+\rho_{24}-\rho_{12}-\rho_{34})\ + \ \sqrt{K_1\,+\,K_2}\,\rho_{34} \big] }  \ ,
\end{equation}

with energy

\begin{equation}\label{Ee4}
\tilde E^{(e)}_0 \  = \  \frac{d}{\sqrt{2\,m}}\,(\sqrt{K_2} \ + \ \sqrt{K_1\,+\,K_2}) \ + \ \frac{K_2}{2}\,\rho_{12}\ ,
\end{equation}

where ${\cal N}^{(e)}={\cal N}^{(e)}(K_1,K_2,m,d)$ in (\ref{Pe4}) is a normalization factor. It is such that the $L^2$-condition $\int {\big(\tilde \psi^{(e)}_0\big)}^2\,d\mathbf{r}_{{}_3}d\mathbf{r}_{{}_4} =1$ holds, where the integration is over the electronic variables $\mathbf{r}_{{}_3}$ and $\mathbf{r}_{{}_4}$ only.

Expression (\ref{Ee4}) is the one we must add to the Hamiltonian of the relative motion of the two heavy particles $m_1=m_2=1$, namely $(-4\,\rho_{12}\,\pa_{\rho_{12}}^2 -
 2\,d\, \pa_{\rho_{12}} + \frac{1}{4}\,\rho_{12})$, to get the nuclear Hamiltonian, i.e.

\begin{equation}\label{Hnuclear4}
  \tilde{\cal H}^{\rm (nuclear)}\ = \ \bigg[-4\,\rho_{12}\,\pa_{\rho_{12}}^2 -
 2\,d\, \pa_{\rho_{12}} + \frac{1}{4}\,\rho_{12} \bigg]\ + \   \frac{d}{\sqrt{2\,m}}\,(\sqrt{K_2} \ + \ \sqrt{K_1\,+\,K_2}) \ + \ \frac{K_2}{2}\,\rho_{12} \ ,
\end{equation}

The ground state 
\[ 
\tilde \psi_0^{(n)} \ \propto \ e^{-\frac{\sqrt{ 1\,+\,2\,K_2}}{4 }\,\rho_{12}} \ ,
\]
of (\ref{Hnuclear4}) is the one of zero quanta with frequency $\sqrt{1\,+\,2\,K_2}$. Hence, the total wave function in the BOA takes the form

\begin{equation}\label{PBOP4}
\tilde \psi^{\rm (BO)}_0  \  = \  {\cal N}^{\rm (BO)} \ e^{ -\frac{1}{4} \, \sqrt{1\,+\,2\,K_2}  \,\rho_{12}\ -\ \frac{1}{2} \sqrt{\frac{m}{2}}\big[\,\sqrt{K_2}\,(\rho_{13}+\rho_{14}+\rho_{23}+\rho_{24}-\rho_{12}-\rho_{34})\ + \ \sqrt{K_1\,+\,K_2}\,\rho_{34} \big] } \ ,
\end{equation}

where ${\cal N}^{\rm (BO)}$ is a normalization constant, and the corresponding energy is given by

\begin{equation}\label{E0B04}
\tilde E^{\rm (BO)}_0  \ = \   \frac{d}{2}\,\bigg(\ \sqrt{(1\,+2\,K_2)}\  + \  \sqrt{\frac{2\,K_2}{m}} \ + \ \sqrt{\frac{2\,(K_1\,+\,K_2)}{m}} \ \bigg)  \ .
\end{equation}

\subsubsection{Accuracy of the Born-Oppenheimer approximation}

To estimate the accuracy of the BOA, we compute the ratio of the exact and approximate energies in powers of the small mass $m\ll1$, namely

\begin{equation}\label{ratio4}
\begin{aligned}
\Delta \tilde E\ \equiv \ \frac{\tilde{E}_0 \ - \ \tilde E^{\rm (BO)}_0}{\tilde{E}_0}  \ & = \ 1 \ - \ \frac{  \, \sqrt{(1\,+2\,K_2)}\  + \  \sqrt{\frac{2\,K_2}{m}} \ + \ \sqrt{\frac{2\,(K_1\,+\,K_2)}{m}} \ }{ \, \sqrt{1+2\,K_2} \ +\ \sqrt{\frac{2\,(K_1+K_2)}{m}} \ +\ \sqrt{\frac{2\,K_2\,(1+m)}{m}} }
\\ &
\approx \frac{1}{2}\frac{\sqrt{K_2}}{\sqrt{K_2}+\sqrt{K_1+K_2}}\,m \ - \ \frac{1}{2\sqrt{2}}\frac{\sqrt{(1+2K_2)K_2}}{(\sqrt{K_2}+\sqrt{K_1+K_2})^2}\,m^{3/2} \ \dots  \ .
\end{aligned}
\end{equation}

If (i) $K_1$ and $K_2$ are of the order of $1$, which implies that the strength of the interaction between
light and heavy particles is of the same order as between the light
particles and (ii) $m = 1/2000$, which is approximately the relation between the electron and
proton masses, then the ratio (\ref{ratio4}) is $1.024 \times 10^{-4}$. It means that the BOA for the four-body system we study is slightly more accurate than for the three-body case. In the case of the chemical compound $H_2O_2$ (Hydrogen peroxide), $m\cong 1/15$, the fractional energy correction is approximately $0.012\, $, see Fig \ref{P1}.

\begin{figure}[h]
  \centering
  \includegraphics[width=7.0cm]{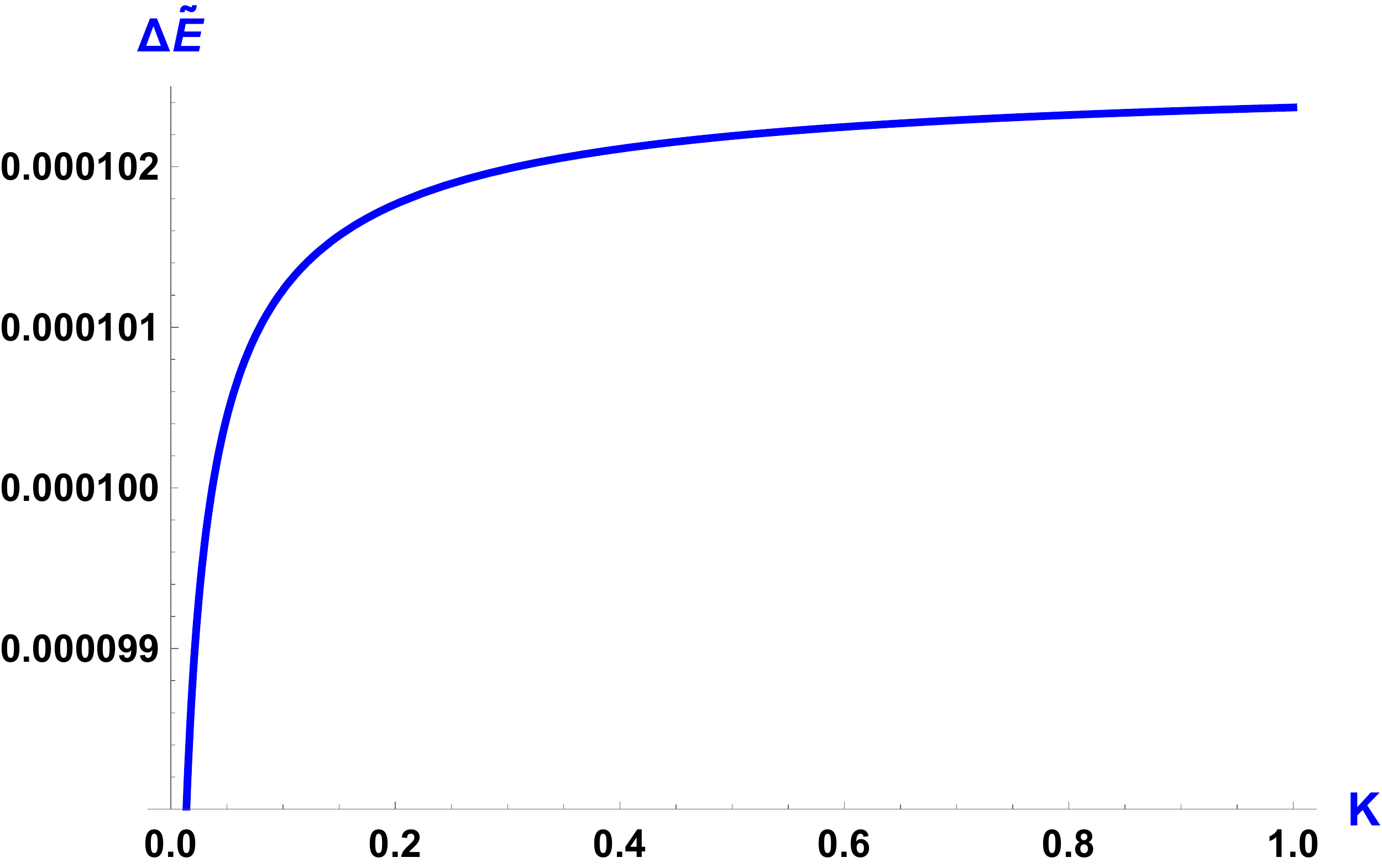} \qquad  \includegraphics[width=7.0cm]{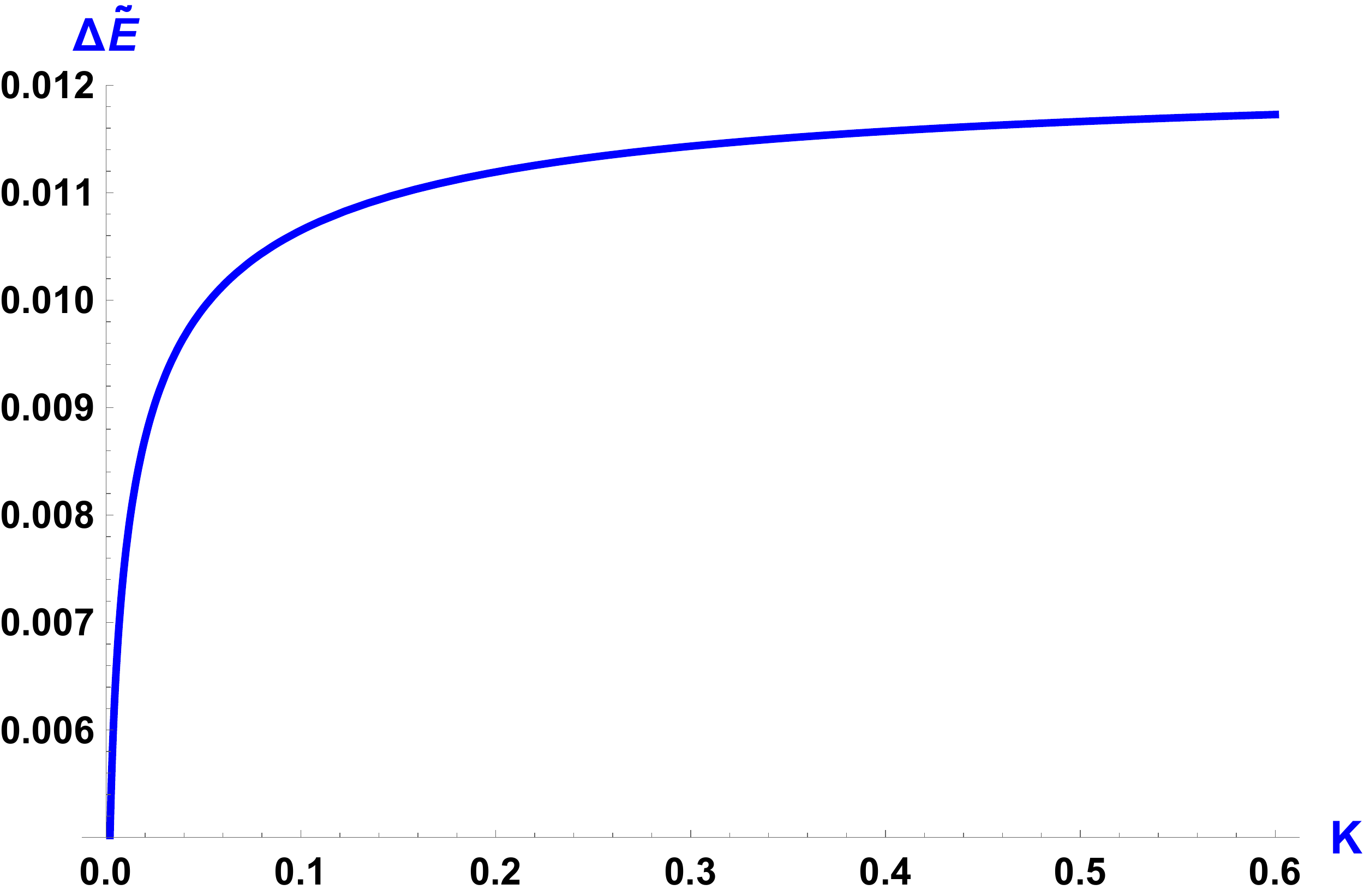}
  \caption{The ratio $\Delta \tilde E$ (\ref{ratio4}) as a function of the spring constant $K\equiv K_1=K_2$ for (left) the $H_2$ molecule ($m=\frac{1}{2000}$) and the chemical compound $H_2O_2$ ($m=\frac{1}{15}$). The ratio $\Delta \tilde E$ is a monotonic increasing (bounded) function of $K$.}
  \label{P1}
\end{figure}

The ratio $\Delta \tilde E$ does not depend on the dimension $d$.  As a function of the light mass $m$, the expansion of the exact energy (\ref{E04}) is given by

\begin{equation}\label{E0B0}
\tilde{E}_0  \ = \   \tilde E^{\rm (BO)}_0 \ + \ \frac{d\,\sqrt{K_2}}{2\,\sqrt{2}}\,\bigg( \,m^{\frac{1}{2}}\ - \   \frac{1}{4}\,m^{\frac{3}{2}} \ + \   \frac{1}{8}\,m^{\frac{5}{2}} \ + \ \dots \bigg)  \ ,
\end{equation}

hence, again the sum of the first two terms correspond to the energy (\ref{E0B04}) obtained in the BOA. Denoting the phases in (\ref{P04}}) and (\ref{PBOP4}) as $\tilde \Phi$ and $\tilde \Phi_{\rm BO}$, respectively, we easily obtain the relation

\begin{equation}\label{}
\tilde \Phi  \ = \  \tilde \Phi_{\rm BO} \ + \ \frac{1}{4}\sqrt{\frac{K_2}{2}} \bigg(\,m^{\frac{3}{2}} \ -\ \frac{3}{4}\,m^{\frac{5}{2}} \ + \ {\cal O}(\,m^{\frac{7}{2}}) \bigg) (\rho_{13}+\rho_{14}+\rho_{23}+\rho_{24}-\rho_{12}-\rho_{34}) \ ,
\end{equation}

that tell us that the BOA provides the lowest terms of the Puiseux series of the exact result.

\section{Many-body system}

In this section, we generalize the three-body $(n=3)$ and four-body $(n=4)$ systems to the case of an arbitrary number of particles $n$. Assuming $d\geq (n-1)$ and that the potential $V=V_n(\rho)$ solely depends on the $\frac{n(n-1)}{2}$ $\rho$-variables
\[
    \rho_{ij}\ = \ r_{ij}^2\   \qquad \quad i,j=0,1,2,\dots,n \ , \qquad i\,<\,j \ ,
\]

we arrive, eventually, at the $\frac{n(n-1)}{2}$-dimensional eigenvalue problem 
\begin{equation}
\label{H-4-body}
       \,[-\De^{(n)}_{\rm rad}(\rho) \ + \ V_n(\rho) \,]\,\psi(\rho)\ =\ E\, \psi(\rho)\ ,
\end{equation}
in the $\rho$-space of relative motion, where

\begin{equation}
\begin{aligned}
\De^{(n)}_{\rm rad}(\rho) \ & = \
2\,\sum_{i \neq j,i\neq k, j< k}^{n}\,\frac{1}{m_i}(\rho_{ij} \ + \ \rho_{ik} - \rho_{jk})\pa_{\rho_{ij}}\pa_{\rho_{ik}} \ + \
2\,\sum_{i<j}^n \bigg(\frac{m_i+m_j}{m_i m_j} \bigg) \rho_{ij} \pa^2_{\rho_{ij}}
\\ &
\ + \ d\, \sum_{i<j}^n\,\bigg(\frac{m_i+m_j}{m_i m_j}\bigg)\pa_{\rho_{ij}} \ ,
\end{aligned}
\label{HRN}
\end{equation}
where $\mu_{ij}$ is as defined before and $\De^{(n)}_{\rm rad}$ plays the role of kinetic radial operator cf.(\ref{addition3-3r-M}). It was conjectured that the operator
\begin{equation}
\label{HradN}
  {\cal H}^{(n)}_{\rm rad}\ \equiv\ -\De^{(n)}_{\rm rad} \ + \  V_n  \ ,
\end{equation}
is equivalent to a $\frac{n(n-1)}{2}$-dimensional radial Schr\"odinger operator, see \cite{TME-N}. As a function of the $\rho$-variables, the operator (\ref{HRN}) is not $S_{\frac{n(n-1)}{2}}$ permutationally-invariant. Nevertheless, it remains $S_n$ invariant under the permutations of the $n$ particles. 

For the Hamiltonian ${\cal H}^{(n)}_{\rm rad}$ (\ref{HradN}), the configuration space is given by ${\cal V}_n\geq0$ where
${\cal V}_n$ is the simplex (volume) of the \emph{polytope of interaction} whose vertices correspond to the positions of the particles. The quantity ${\cal V}_n$ can be written as a Cayley-Menger determinant \cite{Cayley}. The reduced Hamiltonian ${\cal H}^{(n)}_{\rm rad}$ (\ref{HradN}) is essentially self-adjoint with respect to the radial measure
\begin{equation}
\label{normmeasureN}
d\varrho_n \ =\ {({\cal V}_n)}^{d-n}\,\prod_{i,j=1\,;\,i<j}^{n}d\rho_{ij}\ .
\end{equation}

Although the radial Hamiltonian is essentially self-adjoint it is not in the form of a Laplace-Beltrami operator plus potential. For this to be true a further gauge transformation is needed, see \cite{TME-N}.

\section{Many-body harmonic oscillator}

For the $n$-body problem in $d$ dimensions ($d\geq n-1$), we consider the harmonic potential
\begin{equation}
\label{Vn-es}
   V_n^{(es)}\ =\ 2\,\om^2\,\sum_{i<j}^{n} \nu_{i,j}\,\rho_{ij}  \
\end{equation}
where $\om$ and the $\nu$'s are positive constants. In this case, the reduced Hamiltonian (\ref{HradN})
\begin{equation}
\label{Hnes}
   \tilde {\cal H}^{(es)}_{n,{\rm rad}} \ = \ -\De^{(n)}_{\rm rad} \ + \ V_n^{(es)}  \ ,
\end{equation}
 
is an exactly solvable operator. The ground state function takes the following form
\begin{equation}
\label{Psi0n}
   \Psi_{n,0}^{(es)}\ =\ {\cal N}\,e^{-\om\, (\sum_{i<j}^{n} a_{ij}\,\mu_{ij}\,\rho_{ij})}\ ,
\end{equation}
where ${\cal N}$ is a normalization factor and the parameters $a_{ij}$ in the exponent are related to those of the potential ($\ref{Vn-es}$) through the $\frac{n(n-1)}{2}$ algebraic equations
\begin{equation}
\label{freq-n}
\begin{aligned}
 a_{ij} \ & = \  a_{ij}^2\, \mu _{ij} \ + \ a_{ij}\,\bigg(\sum_{i<k;\,k \neq j}^{n}\frac{1}{m_i}a_{ik} +\sum_{k<j;\,k \neq i}^{n}\frac{1}{m_j}a_{kj} \bigg)
 \\ &
\ - \ \sum_{i<k<j}^{n}\frac{1}{m_k}a_{ik}a_{kj}\ - \ \sum_{i,j<k}^{n}\frac{1}{m_k}a_{ik}a_{jk}  \ - \ \sum_{k<i,j}^{n}\frac{1}{m_k}a_{ki}a_{kj} \ - \ \sum_{j<k<i}^{n}\frac{1}{m_k}a_{ki}a_{jk}
\ .
\end{aligned}
\end{equation}

The ground state energy is given by
\begin{equation}
\label{E00n}
E_{n,0}^{(es)}\ = \ \om \,d\,\sum_{i<j}^{n}a_{i,j} \ .
\end{equation}

\subsection{Case of two equal massive particles}

\subsubsection{Exact result}

Now, we focus on the special case where two particles of equal mass ($m_1=m_2=1$) interact between themselves and with ($n-2$) identical particles ($m_3=m_4=\ldots=m_n=m$) through an harmonic oscillator potential, namely

\begin{equation}
\label{Vn}
   V \ =\    \frac{1}{4}\,\rho_{12} \ + \  \frac{K_{2}}{2}\bigg(\sum_{j=2}^{n}\,\rho_{1j} \ + \ \sum_{j=3}^{n}\,\rho_{2j}\bigg)\ + \ \frac{K_{1}}{2}\sum_{i,j=3\,;i\neq j}^{n}\,\rho_{ij} \ .
\end{equation}

For the Hamiltonian (\ref{Hnes}), the exact ground state energy and its eigenfunction are given by

\begin{equation}\label{E0N}
\begin{aligned}
E_{n,0} \ & = \ d\,\bigg[\alpha \, +\, 2\,(n-2)\,\beta \, +\,\frac{1}{2}(n(n-5)+6)\, \gamma\bigg]
\\ &
 = \ \frac{1}{2} \,d\,\left[ \sqrt{1\,+\,(n-2)\,K_2} \ +\ (n-3)\sqrt{\frac{2\,K_2\,+\,(n-2)\,K_1}{m}} \ +\ \sqrt{\frac{K_2\,(2\,+\,(n-2)\,m)}{m}} \right] \ ,
\end{aligned}
\end{equation}

\begin{equation}\label{psi0n}
\psi_{n,0} \ = \ {\cal N}\, e^{\,-\big(\alpha_n\,\frac{1}{2}\,\rho_{12} \ + \ \beta_n\,\frac{ m}{(m+1)}\,[\sum_{j=2}^{n}\,\rho_{1j} \ + \ \sum_{j=3}^{n}\,\rho_{2j}] \ + \ \gamma_n\,\frac{m}{2}\,\sum_{i,j=3\,;i\neq j}^{n}\,\rho_{ij} \,\big)  } \ ,
\end{equation}

respectively, where

\begin{equation}\label{}
\begin{aligned}
& \alpha_n \ = \ \frac{1}{2} \bigg[\,\sqrt{1\,+\,(n-2)\,K_2} \ - \ (n-2)\sqrt{\frac{K_2\,m}{2\,+\,(n-2)\,m}} \, \bigg] \ ,
\\ &
 \beta_n \ = \ \frac{1}{2}\,\frac{m+1}{m}\,\sqrt{\frac{K_2\,m}{2\,+\,(n-2)\,m}} \ ,
 \\ &
 \gamma_n \ = \ \frac{1}{(n-2)\sqrt{m}}\bigg(  \sqrt{(n-2)\,K_1\, +2\,K_2} \ - \ \sqrt{\frac{4\,K_2}{2\,+\,(n-2)\,m}}  \bigg) \ ,
\end{aligned}
\end{equation}

and ${\cal N}$ is a constant of normalization with respect to the radial measure $d\varrho_n$ (\ref{normmeasureN}).

\subsubsection{Accuracy of the Born-Oppenheimer approximation}

As for the Born-Oppenheimer approximation, we assume that the mass of two particles are equal $m_1=m_2=1$ and much heavier than the remaining ($n-2$) particles which also have the same mass $m\equiv m_3=m_4=\ldots=m_n$. As a function of the light mass $m$, the sum of the first two terms of the expansion of the exact energy (\ref{E0N})

\begin{equation}\label{}
E_{n,0}  \ = \   E^{\rm (BO)}_{n,0} \ + \  \frac{d \,\sqrt{K_2} \,(n-2) }{128\, \sqrt{2}}\bigg(\,32\,m^{\frac{1}{2}}\,-\,4\, m^{\frac{3}{2}}\, (n-2)\,+\,m^{\frac{5}{2}} (n-2)^2 \ +  \ldots\,\bigg)   \ ,
\end{equation}

coincide with the energy $E^{\rm (BO)}_{n,0}$ obtained in the BOA. For the phase
\[
\Phi_n \ \equiv \ -\bigg(\alpha_n\,\frac{1}{2}\,\rho_{12} \ + \ \beta_n\,\frac{ m}{(m+1)}\,\bigg[\sum_{j=2}^{n}\,\rho_{1j} \ + \ \sum_{j=3}^{n}\,\rho_{2j}\bigg] \ + \ \gamma_n\,\frac{m}{2}\,\sum_{i,j=3\,;i\neq j}^{n}\,\rho_{ij} \,\bigg) \ ,
\]
of the ground state in (\ref{psi0n}), its Puiseux series expansion
\begin{equation}\label{}
\begin{aligned}
 \Phi_n  \ & = \   \Phi_{n,\rm BO} \ - \ \frac{(n-2)^2\,\sqrt{K_2}}{16\, \sqrt{2}}\bigg(\,  m^{3/2} \ -\ \frac{3\, m^{5/2} \, (n-2)}{8}\,+ \ldots\,\bigg)   \,\rho_{12}
 \\ &
\ + \  \frac{(n-2)\,\sqrt{K_2}}{8 \,\sqrt{2}}\bigg(m^{3/2}\ -\ \frac{3 \, m^{5/2}\, (n-2)}{8}\,+ \ldots\,\bigg)\, \bigg[ \sum_{j=2}^{n}\,\rho_{1j} \ + \ \sum_{j=3}^{n}\,\rho_{2j}\bigg]
\\ &
 \ - \ \frac{\sqrt{K_2} }{4 \sqrt{2}} \bigg(m^{3/2}\ -\ \frac{3\, m^{5/2} (n-2)}{8} \,+ \ldots\,\bigg)\,\sum_{i,j=3\,;i\neq j}^{n}\,\rho_{ij} \ ,
\end{aligned}
\end{equation}

also shows that its lowest terms reproduce the phase we calculated in the BOA. Finally, with $m \ll 1$, we compute the ratio

\begin{equation}\label{RatioN}
\begin{aligned}
\Delta  E_n \ \equiv \ \frac{{E}_{n,0} \ - \ E^{\rm (BO)}_{n,0}}{{E}_{n,0}}  \ & = \ \frac{\sqrt{K_2} \, (n-2)}{2 \,\sqrt{2} \,\left((n-3) \sqrt{K_1\, (n-2)\,+\,2 \,K_2}\,+\,\sqrt{2\,K_2}\right)}\,m
\\ &
 - \ \frac{\sqrt{K_2}\,  (n-2)\, \sqrt{K_2\, (n-2)\,+\,1}}{2 \,\sqrt{2}\, \left((n-3)\, \sqrt{K_1 \,(n-2)\,+\,2 \,K_2}\,+ \,\sqrt{2\,K_2}\,\right){}^2}\,m^{\frac{3}{2}}\ +
 \dots \ \ ,
\end{aligned}
\end{equation}
to estimate the accuracy of the Born-Oppenheimer approximation. The ratio (\ref{RatioN}), at $n=3$ ($K_1=0$) reduces to expression (\ref{delE}) while at $n=4$ coincides with (\ref{ratio4}). Once again, the ratio $\Delta  E$ does not depend on the dimension $d$. The first term, which is the dominant when $m\ll1$, is always positive while the second term is negative. In the limit $n\gg1$ the first and second term tend to $\frac{1}{2}\sqrt{\frac{K_2}{2K_1n}}m$ and $\frac{K_2}{2\sqrt{2K_1}}m^{\frac{3}{2}}$, respectively.

\section*{Conclusions}

In this paper we studied the quantum system of three particles coupled to each other by harmonic forces. We re-derived and extended to the $d$-dimensional case, using the formalism in Ref. \cite{TME3-d}, previous results (see \cite{Moshinski}) such as the energies and eigenfunctions of the ground state. The exact results are compared with those obtained in the Born-Oppenheimer approximation (BOA), showing explicitly with examples the accuracy for the later.   

We also studied the quantum 4-body problem in a $d$-dimensional space, $d>2$, of two particles of equal heavy mass $m_1=m_2=1$ interacting
between themselves and with two light particles $m_3=m_4=m\ll1$ through harmonic oscillator potentials. For the ground state level, this model is solved both exactly and within the framework of the Born-Oppenheimer approximation.

We have shown that the ratio between the energies of the approximate and exact solutions is $d$-independent and differs from unity by terms of the order of the dimensionless ratio $m$ of the masses. For the phase $\Phi$ of the ground state wave function and the corresponding ground state energy $E_0$, the approximate and exact solutions are related. The first terms of the Puiseux series expansion (in powers of $m$) of the exact results coincide exactly with the approximate solutions obtained in the BOA. Two physically relevant examples where considered where the light particles are either electrons ($H_2$ molecule) or protons ($H_2O_2$ compound).

The generalization to an arbitrary number $n$ of particles interacting through an harmonic oscillator potential in a $d-$dimensional space ($d\geq n-1$) is discussed as well. In the case of two particles with equal heavy mass ($m_1=m_2=1$) and ($n-2$) light particles ($m_3=m_4=\ldots=m_n=m \ll 1$), we found that the ratio between the energies of the approximate and exact solutions is again $d$-independent, and differs from unity by a term proportional to the ratio of the masses with an $n$-dependent coefficient that vanishes as $\sim \frac{1}{\sqrt{n}}$ at $n\rightarrow \infty$. We hope that the present consideration can be exploited and lead to approaches much better than the Born-Oppenheimer approximation.

\acknowledgements

The authors are grateful to A. M. Escobar-Ruiz who initiated this work and gave important remarks during its realization. A. M.-R. acknowledges support from DGAPA-UNAM Project No. IA101320. C. A. E. is supported by a UNAM- DGAPA postdoctoral fellowship and Project PAPIIT No. IN111518.

\end{document}